\documentclass[aps,pra,paper,reprint,amsmath,amssymb,groupedaddress,floatfix,showpacs,showkeys,subeqn]{revtex4-1} 
\usepackage{hyperref}
\usepackage{graphics} 
\usepackage{graphicx} 
\usepackage{subfigure}
\usepackage{url}
\usepackage{tipa}

\newcommand{\sech}{\, \mathrm{sech} \,}
\newcommand{\field}[1]{\mathbb{#1}} % requires amsfonts
\begin{document} 
\author{L. A. Toikka}
\author{O. K\"{a}rki}
\author{K.-A. Suominen}
\affiliation{Turku Centre for Quantum Physics, Department of Physics and Astronomy, University of Turku, 20014 Turku, Finland}
\title{Creation and revival of ring dark solitons in a toroidal Bose-Einstein condensate}
\date{\today}
\begin{abstract}
We propose a protocol for the simultaneous controlled creation of multiple concentric ring dark solitons in a toroidally trapped Bose-Einstein condensate. The decay of these solitons into a vortex-antivortex necklace shows revivals of the soliton structure, but eventually becomes an example of quantum turbulence. 
\end{abstract}

\maketitle

The nonlinear nature of a Bose-Einstein condensate (BEC) makes it possible to support dark solitons~\cite{Kivshar1998,P&S,Emergent,Greekreview}, which display a remarkable form-stability, e.g. they propagate without dispersion~\cite{Drazin1989}. This stability is possible because the nonlinearity of the system has an opposite and cancelling effect to dispersion, but the stability can be lost, in general, when the dark solitons decay into vortex-antivortex pairs via the snake instability~\cite{PhysRevA.60.R2665,PhysRevA.64.063602,ZacharyDutton07272001, PhysRevLett.86.2926,PhysRevA.87.043601}.

Dark solitons can be experimentally prepared by creating a phase step in the condensate and letting the density respond by generating the dark soliton (phase imprinting)~\cite{PhysRevLett.83.5198,J.Denschlag01072000,Becker08}, but it is also possible to consider the complementary process of imposing a density profile and letting the phase respond by producing a step (density imprinting). In Refs.~\cite{0953-4075-30-22-001,Shomroni09}, density imprinting is achieved by letting two BECs collide and produce an interference pattern, which then evolves into dark (or grey) solitons. 

In this paper, we consider the challenging case of ring dark solitons (RDSs), which have not yet been observed with cold atoms. Compared to planar solitons with $\tanh$-shaped amplitude, analytic descriptions for RDSs are not easy to obtain~\cite{ths2012}. Instead of considering the interference of two condensates~\cite{Andrews31011997}, we propose a density imprinting protocol involving a time-dependent double-well potential to make a single condensate produce fringes, which are let to evolve into ring dark solitons. We show how the fringes could arise from self-interference, which underlines the long-range coherence of a BEC, and is an example of nonlinear matter-wave interference~\cite{0953-4075-31-8-001}. In Ref.~\cite{PhysRevA.74.043613}, RDSs were defined by  nodes of numerically found radial solutions, resembling the regular Bessel functions in the noninteracting limit. This approach is similar to our case, where the RDSs also evolve from nodes in the wavefunction.

Our idea for the protocol arises from the APLIP (adiabatic passage by light-induced potentials) method for molecular bond extension~\cite{Garraway1998,Rodriguez2000,PROP:PROP200310015}. The method relies on three states with spatially dependent potentials, which are coupled by laser pulses that are applied in counterintuitive order. It is similar to the widely used method of stimulated Raman adiabatic passage (STIRAP)~\cite{vitanov2001laser}, as used in Refs.~\cite{Ljp47318,PhysRevLett.80.2972}, for example, but with important differences: e.g. it can be reduced to a single-state process~\cite{Harkonen2006} in two dimensions. The success of APLIP in the noninteracting case makes it plausible to extend similar ideas also to the weakly-interacting BEC.

Focussing on the case of cylindrical symmetry as expected for ring dark solitons, we present an analytical expression for the time-dependent double-well potential that produces from one to four concentric RDSs. We note that similar potentials would work for the creation of planar solitons as well~\cite{Harkonen2006}. The stability of the RDS is discussed, and we show that the original RDS(s) decay into vortex-antivortex necklaces when the snake instability is induced, but remarkably show revivals before the ultimate (and spontaneous) onset of quantum turbulence. 

To begin, let us consider a scalar order parameter, $\psi$, representing the macroscopic wavefunction of a Bose-Einstein condensate trapped in a potential given by $V_{\mathrm{trap}}$, which is a solution to the Gross-Pitaevskii equation:
\begin{equation}
\label{eqn:nls0}
i\psi_t = -\nabla^2 \psi + V_{\mathrm{trap}}\psi + C_{\mathrm{2D}}|\psi^2|\psi.
\end{equation}
Here we have assumed a two-dimensional condensate whereby the $z$-direction is tightly trapped to the corresponding harmonic oscillator ground state ($\omega_z \gg \omega_x = \omega_y \equiv \omega_r$) and has been projected onto the $xy$-plane. Then $C_{\mathrm{2D}} = 4 \sqrt{\pi} Na/a_{\text{osc}}^{(z)}$, where $N$, $a$, and $a_{\text{osc}}^{(z)}$ are the number of atoms in the cloud, the $s$-wave scattering length of the atoms, and the characteristic trap length in the $z$-direction respectively. We have obtained dimensionless quantities by measuring time, length and energy in terms of $\omega_r^{-1}$, $a_{\text{osc}}^{(r)} \equiv a_{\text{osc}} = \sqrt{\hbar/(2m\omega_r})$ and $\hbar \omega_r$ respectively, where $\omega_r$ is the angular frequency of the trap in the $r$-direction. This basis is equivalent to setting $\omega_r = \hbar = 2m = 1$.

\begin{figure*}[ht]
\centering
\includegraphics[width=\textwidth] {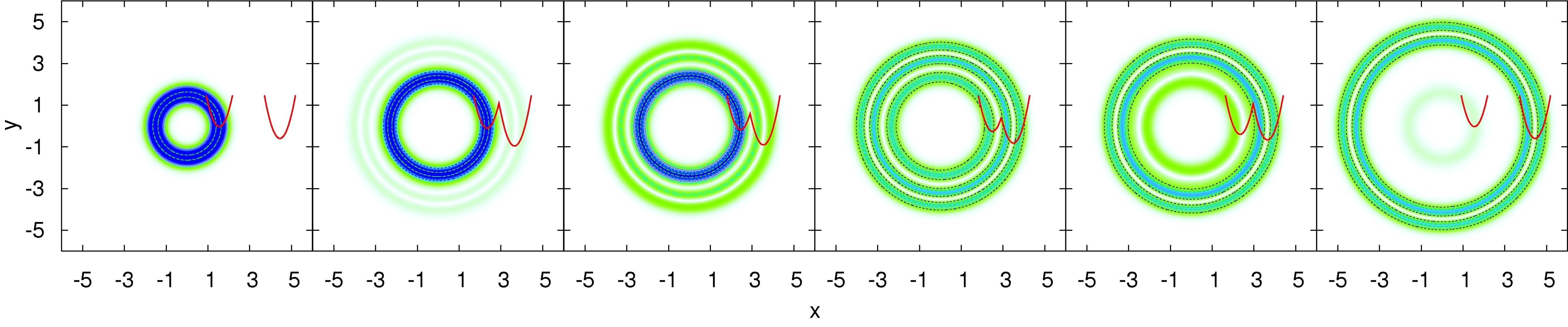}
\caption{\label{fig:APLIP_V_APPROX_XY} (Color online.) Creation of one ring dark soliton in a toroidal BEC using a time-dependent double-well potential (see  $\tilde{V}$ in Eq.~\eqref{eqn:APLIP_V_APPROX}) (red solid line), with $\Delta_2 = -15.0$. As soon as the outer well becomes accessible, the condensate produces a fringe that then evolves into the RDS. The simulation was performed on an $(x,y)$ grid, which has been shown to be enough to induce the snake instability by breaking the cylindrical symmetry~\cite{PhysRevA.87.043601}, but in the time scales, and with the value of $C_{\mathrm{2D}} = 50$ relevant for the creation process here, the ring soliton remains stable. We have numerically confirmed that the stability remains if we use (slightly) larger radii and/or smaller nonlinearities, but a higher $C_{\mathrm{2D}}$ will induce the instability, as discussed in the main text. From left to right, the times are $0.0, 10.5, 11.5, 12.5, 14.5$, and $25.0$. }
\end{figure*}

By choosing $V_{\mathrm{trap}}$ appropriately, it is possible to control the quantum dynamics of the condensate. The main result of this paper is a simple analytic time dependence for $V_{\mathrm{trap}}$ that generates one or multiple concentric ring dark solitons. We will not perform an exhaustive mapping of the large parameter space, but the process can be extended to an even higher number of solitons.

To generate multiple concentric ring dark solitons, $V_{\mathrm{trap}}$ takes the form of a combination of two simple harmonic potentials with time-dependent energy minima and radii, i.e. we set $V_{\mathrm{trap}} = \tilde{V}$, where
\begin{equation}
\label{eqn:APLIP_V_APPROX}
\tilde{V}(r,t) = \begin{cases} \tilde{V}_1(r,t) &\mbox{if } r \leq r_v \\
\tilde{V}_2(r,t) & \mbox{if } r > r_v, \end{cases} 
\end{equation}
where ($i = 1,2$):
\begin{align}
\tilde{V}_i(r,t) &= \frac{1}{4}\omega^2 (r - \alpha_i - \beta_i \sech{[\gamma_i(t-t_s)]})^2 + \Delta_i(t), \\
\Delta_i (t) &= \left(30 -\frac{2}{5}\Omega \right) e^{-(t-t_i)^2/(T-\gamma_i(t_1-t_2))^2} + \Delta_i, \notag
\end{align}
and where $r_v$ is the radius of the point of intersection of $\tilde{V}_{1,2}$. As initial state we choose the ground state of the potential $\tilde{V}_1$ at $t = 0$; for the parameters we use $\omega = 20.0, \Delta_1 = 0.0, t_1 = 15.0, t_2 = 10.0, T = 5.0, t_s = 12.5, \alpha_1 = 1.5, \beta_1 = -\beta_2 = 1.0, \gamma_1 = \gamma_2 = 0.30$, and $\alpha_2 = 4.5$. Throughout this paper we use $C_{\mathrm{2D}} = 50.0$ unless stated otherwise. The remaining parameters $\Delta_2$ and $\Omega \in [100, 175]$ determine the number of ring dark solitons produced. Without loss of generality, we set $\Omega = 100.0$, and $\Delta_2$ is separately specified for each case. We note that for $\Omega \gtrsim 175$, or for different values of the other parameters, Eq.~\eqref{eqn:APLIP_V_APPROX} might become less accurate. Furthermore, based on the mapping of the $(\Delta_2, \Omega )$ space performed in a previous study considering the linear ($C_{\mathrm{2D}} = 0$) case~\cite{Harkonen2006}, we expect there to be certain values of $\Omega$ (e.g. $\Omega \approx 145.0$) for which this process is not expected to work either in the interacting BEC case.

\begin{figure}
\centering
\includegraphics[width=0.4\textwidth] {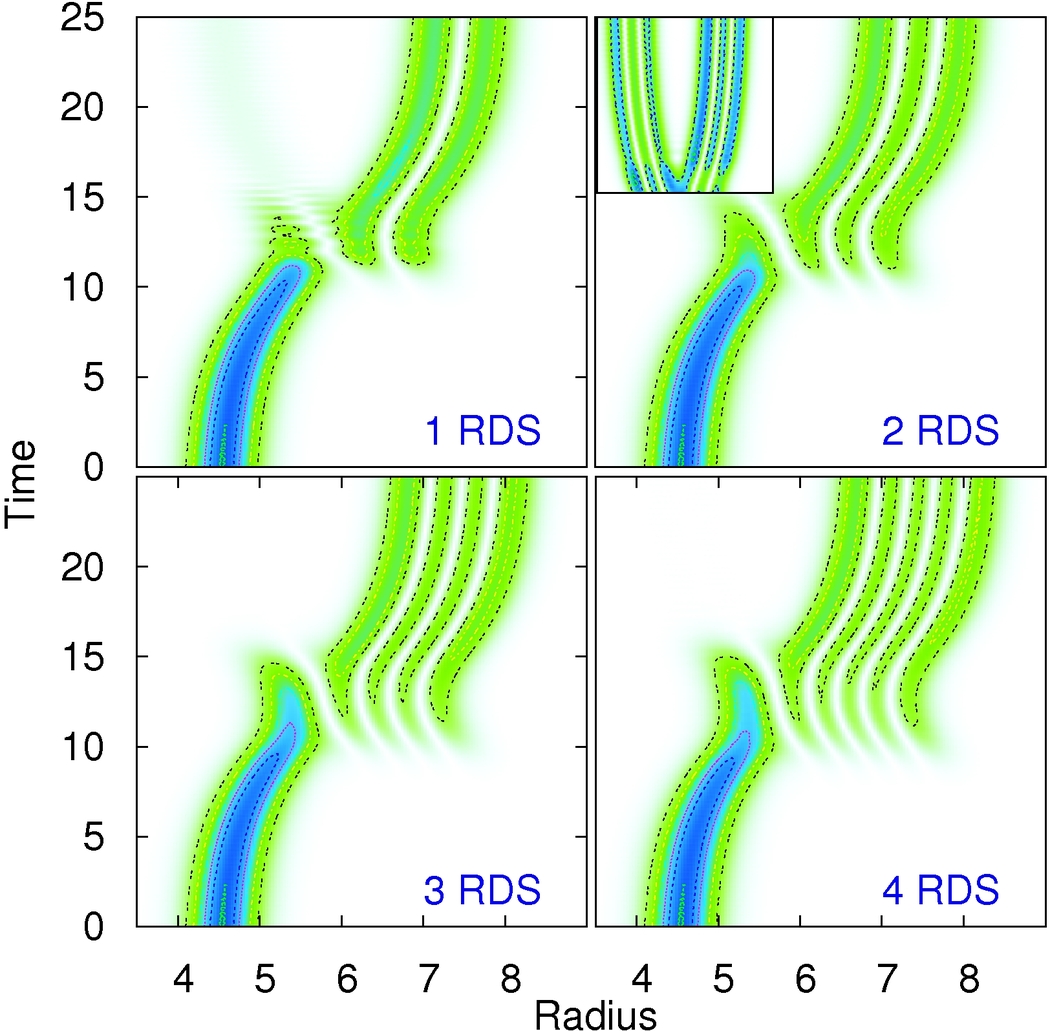}
\caption{\label{fig:APLIP_V_APPROX} (Color online.) Density plots of the creation of multiple ring dark solitons (RDS) in a toroidal BEC using the potential $\tilde{V}$, with $\alpha_1 = 4.5$, and $\alpha_2 = 7.5$. (1RDS) $\Delta_2 = -15.0$. (2RDS) $\Delta_2 = -35.0$. (3RDS) $\Delta_2 = -60.0$. (4RDS) $\Delta_2 = -80.0$. The inset in 2RDS shows the evolution for $\Omega = 175.0$: the potential surface produces two ring dark solitons in both wells. See Ref.~\cite{SM_Video} for a video of the creation process (in the video the potentials are slightly softened around $r_v$, but we still achieve the state transfer with high fidelity).}
\end{figure}

The time evolution given by Eqs.~\eqref{eqn:nls0} and~\eqref{eqn:APLIP_V_APPROX} is shown in Figs.~\ref{fig:APLIP_V_APPROX_XY} and~\ref{fig:APLIP_V_APPROX}. In Fig.~\ref{fig:APLIP_V_APPROX_XY}, we produce a single ring dark soliton, which remains stable throughout the process because we use a sufficiently low value of $C_{\mathrm{2D}}$ to suppress the decay into vortex-antivortex pairs via the snake instability~\cite{PhysRevA.87.043601}. With various choices for $\Delta_2$, we obtain between one and four ring dark solitons (see Fig.~\ref{fig:APLIP_V_APPROX}). Note that we could equally well produce the soliton structure so that it occupies the inner potential well at the end of the process~\cite{Rodriguez2000,Harkonen2006}.

The process in general relies on the adiabatic passage of the initial state with possible diabatic jumps, but the visible changes are due to the adiabatic state itself changing. Our explanation for the mechanism is that the condensate in the outer well undergoes reflection and self-interference; subsequently, the ring dark solitons evolve from the fringes. This process can be modelled either by the method of images~\cite{Toikka2013} or simply by the de-Broglie wavelength. Both ways give the approximate fringe spacing of $\Delta r \approx \pi/k$, where $k$ is the momentum imparted by the outer well. For $k = \sqrt{|\Delta_2|}$ with $\Delta_2 = -n\omega$, where $n \in \field{N}^+$, we therefore obtain that the total number of fringes in the outer well is $4n/\pi \approx 1.27 n$, which rounding down to the nearest integer is simply $n$, agreeing with the choices of $\Delta_2$ in Fig.~\ref{fig:APLIP_V_APPROX}.

The time evolution is quite sensitive to the form of the potential around $r_v$, with the height of the barrier between the two wells playing a crucial role. It is important that this barrier be gradually lowered, preventing full sudden expansion of the inner well condensate onto the outer well to ensure smooth control. However, the process is robust against small variations of the parameters and for small nonlinearities ($C_{\mathrm{2D}} \lesssim 50$), beyond which there can be significant residue left in the inner well because of loss of adiabaticity, and even total loss of the ring soliton(s) via the snake instability. For higher nonlinearities, one can use a Feshbach resonance~\cite{PhysRevLett.102.090402} to tune $C_{\mathrm{2D}}$ to be small enough for the process described here to work with high fidelity, and then ramp it back up after the population transfer.

\begin{figure}
\centering
\includegraphics[width=0.45\textwidth] {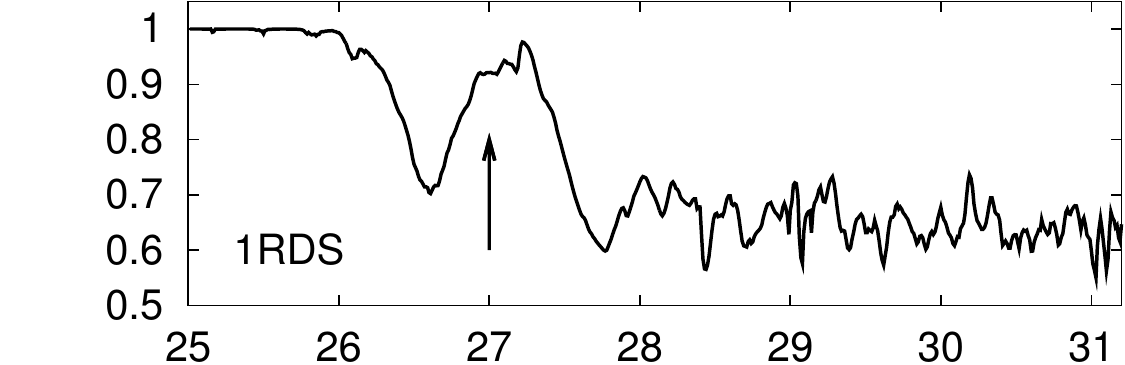}
\includegraphics[width=0.45\textwidth] {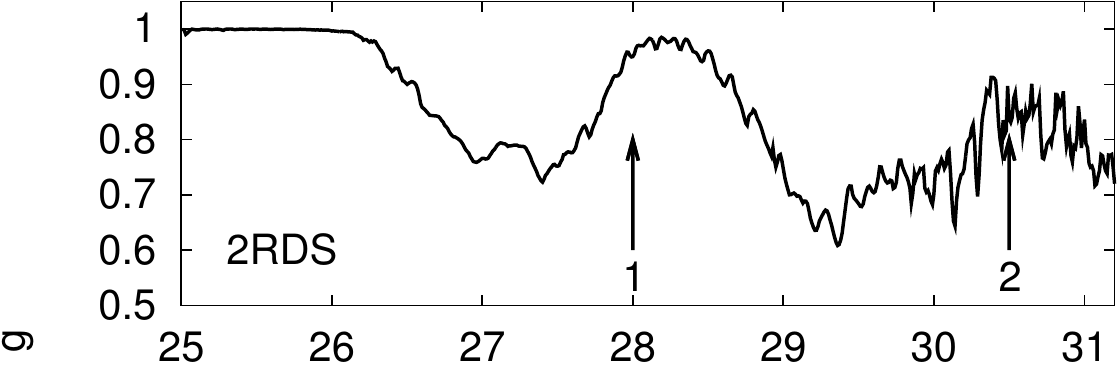}
\includegraphics[width=0.45\textwidth] {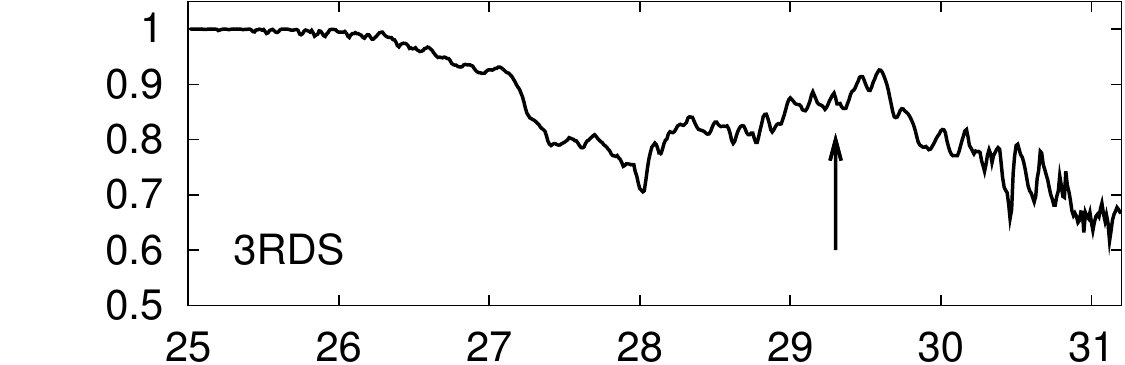}
\includegraphics[width=0.45\textwidth] {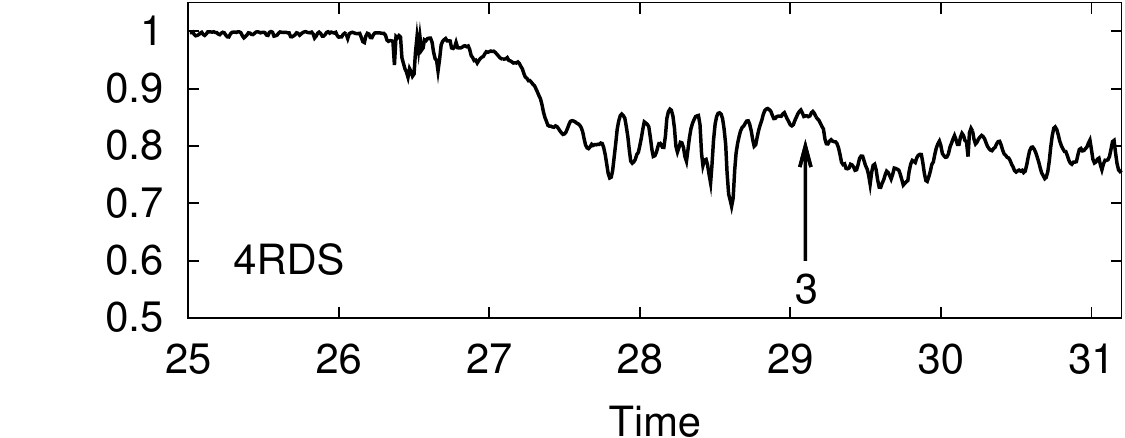}
\includegraphics[width=0.5\textwidth] {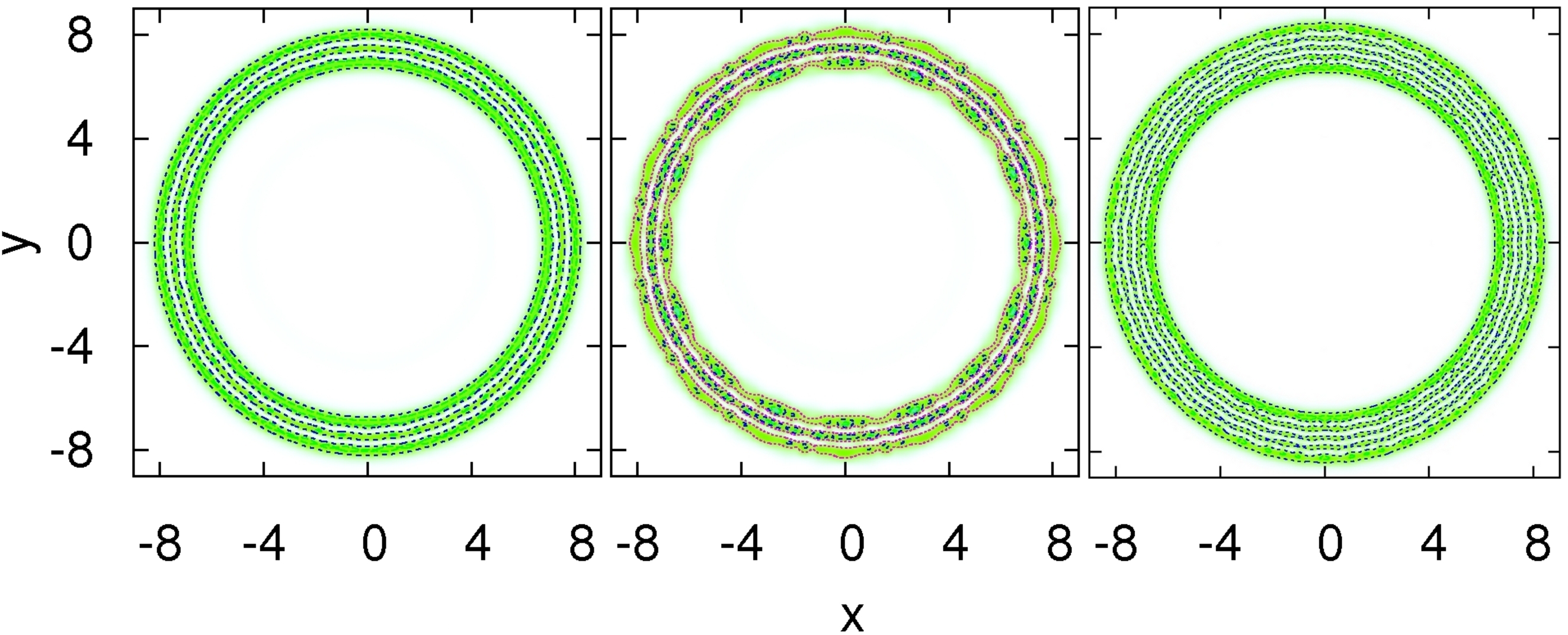}
\caption{\label{fig:g1all} (Color online.) The modified correlation function $g$ (see Eq.~\eqref{eqn:angcorrfunc1}) as a function of time after the RDSs have been created and the nonlinearity has been set to $C_{\mathrm{2D}} = 400$, which induces the snake instability. Here $\alpha_2 = 4.5$ for 1RDS and $\alpha_2 = 7.5$ otherwise. The revivals of the RDSs can be seen as recoverings of phase coherence, but ultimately the decoherence leads to the onset of quantum turbulence. The arrows indicate when the RDS is revived. The bottom panel shows the density $|\psi|^2$ at the times indicated by the arrows labeled by "1", "2", and "3" from left to right respectively.}
\end{figure}

Higher $C_{\mathrm{2D}}$, however, will induce the snake instability of the RDS. To study the decay, we set nonadiabatically $C_{\mathrm{2D}} = 400$ at $t = 25$. Interestingly, the RDSs are revived after the first snake instability, after which the regime of quantum turbulence begins. This behaviour is similar to an experimentally observed oscillating soliton/vortex ring~\cite{Shomroni09}, but here the soliton is much longer giving rise to more vortex-antivortex pairs. To gain insight on the decay and revival, we consider the normalized equal-time first-order correlation function (see, e.g. Ref.~\cite{gerry2005introductory})
\begin{equation}
\label{eqn:angcorrfunc}
g^{(1)}(\textbf{r}_1,\textbf{r}_2) = \frac{\frac{1}{2} \left\langle \hat{\psi}^\dagger (\textbf{r}_1)\hat{\psi}(\textbf{r}_2) + \hat{\psi}^\dagger (\textbf{r}_2) \hat{\psi}(\textbf{r}_1) \right\rangle}{\sqrt{\left\langle \hat{\psi}^\dagger (\textbf{r}_1)\hat{\psi}(\textbf{r}_1) \right\rangle \left\langle \hat{\psi}^\dagger (\textbf{r}_2)\hat{\psi}(\textbf{r}_2) \right\rangle} },
\end{equation}
for which $0 \leq |g^{(1)}(\textbf{r}_1,\textbf{r}_2)| \leq 1$, and where $\textbf{r}_1 = (r,0)$ and $\textbf{r}_2 = (r,\theta)$. Let us further define a dimensionless quantity, $g$, by
\begin{equation}
\label{eqn:angcorrfunc1}
g = \frac{1}{N_r} \sum_{r = \alpha_2-\epsilon}^{\alpha_2+\epsilon} \left\langle |g^{(1)}(\textbf{r}_1,\textbf{r}_2)| \right\rangle _\theta,
\end{equation}
where we take $r$ equally spaced with $N_r = 10$ steps over the range of $2\epsilon$ (chosen to cover the ring), and where $\langle \bullet \rangle_\theta$ denotes average over $\theta$. Figure~\ref{fig:g1all} shows the dependence of $g$ on time after $t = 25$. Local (the limit of the correlation function at infinity is not necessarily affected) phase coherence is gradually lost, but the vortex-antivortex necklace(s) recombine(s) back to produce RDS(s). As is evident in Fig.~\ref{fig:g1all}, a higher number of ring dark solitons slows down the loss of coherence and delays the onset of the decay and subsequent revival, but the effect seems to saturate above 3RDS. Eventually, the (local) decoherence is enough for the onset of quantum turbulence, preventing orderly dynamics needed by further revivals of the RDS.

\begin{figure}
\centering
\includegraphics[width=0.4\textwidth] {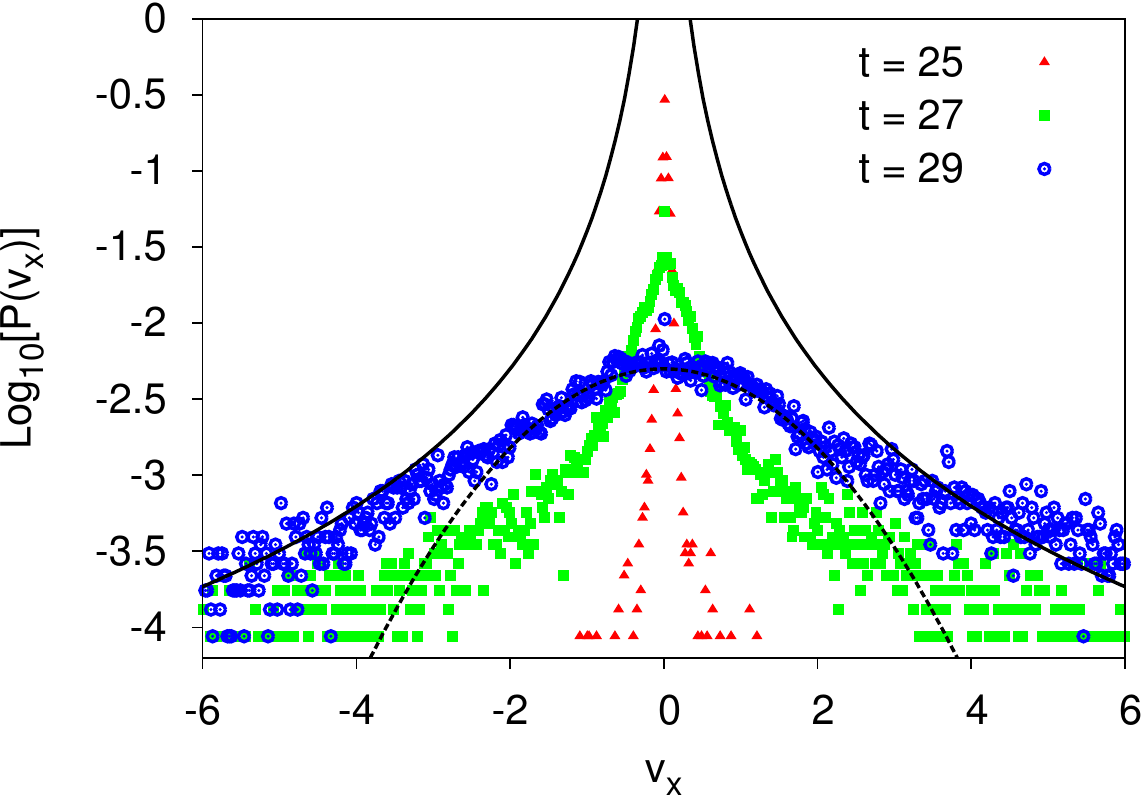}
\caption{\label{fig:QT} (Color online.) Superfluid velocity statistics at three different times for the 1RDS case, showing the emergence of power-law tails characteristic of quantum turbulence. At around $t = 27$, the vortex-antivortex necklace recombines to reproduce the RDS, which then later decays again. The solid line shows $P(v_x) \propto v_x^{-3}$, and the dashed line shows a Gaussian distribution, characteristic of classical turbulence.}
\end{figure}

Quantum turbulence, a complex spaghetti of tangled vortex lines, can be characterised by power-law tails in the superfluid velocity probability distribution $P(v)$ (see Refs.~\cite{PhysRevLett.104.075301,PhysRevE.84.067301} and references therein), making a difference to classical turbulence with Gaussian distributions. This result has been experimentally shown by using solid hydrogen tracers (much smaller than the average vortex separation) in superfluid helium~\cite{PhysRevLett.101.154501}, giving $P(v_i) \propto v_i^{-3}$, where $i$ refers to the velocity component. 

We calculate the velocity distributions $\textbf{v} \propto \nabla S$, where $S$ is the phase (see Fig.~\ref{fig:QT}). We observe the emergence of power-law tails in the velocity statistics as the modified coherence function $g$ becomes smaller, which signifies the onset of quantum turbulence, in accordance with the previous literature.

In summary, we have proposed a novel protocol for the creation of multiple concentric ring dark solitons in a Bose-Einstein condensate. The RDSs evolve from the self-interference fringes, and show revivals upon decay through the snake instability until the regime of quantum turbulence starts and the orderly revival is inhibited. This observation is an example of coherent vortex dynamics before the full onset of quantum turbulence. The results demonstrate a straightforward experimental protocol for the creation of dark solitons and ring dark solitons in scalar BECs, provided that the time-dependence of the trapping potential can be implemented. One possibility is to use time-averaged potentials, see Refs.~\cite{PhysRevLett.80.3899,PhysRevA.64.063602}. 

We acknowledge the support of the Academy of Finland (grant 133682) 
and Jenny and Antti Wihuri Foundation (LT). We also thank B. P. Anderson for fruitful discussions (LT).

\bibliographystyle{apsrev4-1}
\bibliography{../references}

%Merlin.mbs v4.21 2009-07-09.
\begin{thebibliography}{10}%
\makeatletter
\providecommand \@ifxundefined [1]{%
 \ifx #1\undefined \expandafter \@firstoftwo
 \else \expandafter \@secondoftwo
\fi
}%
\providecommand \@ifnum [1]{%
 \ifnum #1\expandafter \@firstoftwo
 \else \expandafter \@secondoftwo
\fi
}%
\providecommand \enquote [1]{``#1''}%
\providecommand \bibnamefont  [1]{#1}%
\providecommand \bibfnamefont [1]{#1}%
\providecommand \citenamefont [1]{#1}%
\providecommand\href[0]{\@sanitize\@href}%
\providecommand\@href[1]{\endgroup\@@startlink{#1}\endgroup\@@href}%
\providecommand\@@href[1]{#1\@@endlink}%
\providecommand \@sanitize [0]{\begingroup\catcode`\&12\catcode`\#12\relax}%
\@ifxundefined \pdfoutput {\@firstoftwo}{%
 \@ifnum{\z@=\pdfoutput}{\@firstoftwo}{\@secondoftwo}%
}{%
 \providecommand\@@startlink[1]{\leavevmode\special{html:<a href="#1">}}%
 \providecommand\@@endlink[0]{\special{html:</a>}}%
}{%
 \providecommand\@@startlink[1]{%
  \leavevmode
  \pdfstartlink
   attr{/Border[0 0 1 ]/H/I/C[0 1 1]}%
   user{/Subtype/Link/A<</Type/Action/S/URI/URI(#1)>>}%
  \relax
 }%
 \providecommand\@@endlink[0]{\pdfendlink}%
}%
\providecommand \url  [0]{\begingroup\@sanitize \@url }%
\providecommand \@url [1]{\endgroup\@href {#1}{\urlprefix}}%
\providecommand \urlprefix [0]{URL }%
\providecommand \Eprint[0]{\href }%
\@ifxundefined \urlstyle {%
  \providecommand \doi [1]{doi:\discretionary{}{}{}#1}%
}{%
  \providecommand \doi [0]{doi:\discretionary{}{}{}\begingroup
  \urlstyle{rm}\Url }%
}%
\providecommand \doibase [0]{http://dx.doi.org/}%
\providecommand \Doi[1]{\href{\doibase#1}}%
\providecommand \bibAnnote [3]{%
  \BibitemShut{#1}%
  \begin{quotation}\noindent
    \textsc{Key:}\ #2\\\textsc{Annotation:}\ #3%
  \end{quotation}%
}%
\providecommand \bibAnnoteFile [2]{%
  \IfFileExists{#2}{\bibAnnote {#1} {#2} {\input{#2}}}{}%
}%
\providecommand \typeout [0]{\immediate \write \m@ne }%
\providecommand \selectlanguage [0]{\@gobble}%
\providecommand \bibinfo [0]{\@secondoftwo}%
\providecommand \bibfield [0]{\@secondoftwo}%
\providecommand \translation [1]{[#1]}%
\providecommand \BibitemOpen[0]{}%
\providecommand \bibitemStop [0]{}%
\providecommand \bibitemNoStop [0]{.\EOS\space}%
\providecommand \EOS [0]{\spacefactor3000\relax}%
\providecommand \BibitemShut [1]{\csname bibitem#1\endcsname}%
%</preamble>
\bibitem{Kivshar1998}%
  \BibitemOpen
  \bibfield{author}{%
  \bibinfo {author} {\bibfnamefont{Y.~S.}\ \bibnamefont{Kivshar}}\ and\
  \bibinfo {author} {\bibfnamefont{B.}~\bibnamefont{Luther-Davies}},\ }%
  \bibfield{journal}{%
  \bibinfo {journal} {Phys. Rep.}\ }%
  \textbf{\bibinfo {volume} {298}},\ \bibinfo {pages} {81} (\bibinfo {year}
  {1998})%
  \bibAnnoteFile{NoStop}{Kivshar1998}%
\bibitem{P&S}%
  \BibitemOpen
  \bibfield{author}{%
  \bibinfo {author} {\bibfnamefont{C.}~\bibnamefont{Pethick}}\ and\ \bibinfo
  {author} {\bibfnamefont{H.}~\bibnamefont{Smith}},\ }%
  \emph{\bibinfo {title} {{B}ose-{E}instein Condensation in Dilute Gases}},\
  \bibinfo {edition} {2nd}\ ed.\ (\bibinfo {publisher} {Cambridge University
  Press},\ \bibinfo {year} {2008})%
  \bibAnnoteFile{NoStop}{P&S}%
\bibitem{Emergent}%
  \BibitemOpen
  \bibfield{author}{%
  \bibinfo {author} {\bibfnamefont{P.~G.}\ \bibnamefont{Kevrekidis}}, \bibinfo
  {author} {\bibfnamefont{D.~J.}\ \bibnamefont{Frantzeskakis}},\ and\ \bibinfo
  {author} {\bibfnamefont{R.}~\bibnamefont{Carretero-Gonz\'alez~(Eds.)}},\ }%
  \emph{\bibinfo {title} {Emergent Nonlinear Phenomena in {B}ose-{E}instein
  Condensates}}\ (\bibinfo {publisher} {Springer},\ \bibinfo {year} {2008})%
  \bibAnnoteFile{NoStop}{Emergent}%
\bibitem{Greekreview}%
  \BibitemOpen
  \bibfield{author}{%
  \bibinfo {author} {\bibfnamefont{D.~J.}\ \bibnamefont{Frantzeskakis}},\ }%
  \bibfield{journal}{%
  \Doi{10.1088/1751-8113/43/21/213001}{\bibinfo {journal} {J. Phys. A}}\ }%
  \textbf{\bibinfo {volume} {43}},\ \bibinfo {pages} {213001} (\bibinfo {year}
  {2010})%
  \bibAnnoteFile{NoStop}{Greekreview}%
\bibitem{Drazin1989}%
  \BibitemOpen
  \bibfield{author}{%
  \bibinfo {author} {\bibfnamefont{P.~G.}\ \bibnamefont{Drazin}}\ and\ \bibinfo
  {author} {\bibfnamefont{R.~S.}\ \bibnamefont{Johnson}},\ }%
  \emph{\bibinfo {title} {Solitons: An Introduction}},\ \bibinfo {edition}
  {2nd}\ ed.\ (\bibinfo {publisher} {Cambridge University Press},\ \bibinfo
  {year} {1989})%
  \bibAnnoteFile{NoStop}{Drazin1989}%
\bibitem{PhysRevA.60.R2665}%
  \BibitemOpen
  \bibfield{author}{%
  \bibinfo {author} {\bibfnamefont{A.~E.}\ \bibnamefont{Muryshev}}, \bibinfo
  {author} {\bibfnamefont{H.~B.}\ \bibnamefont{van Linden van~den Heuvell}},\
  and\ \bibinfo {author} {\bibfnamefont{G.~V.}\ \bibnamefont{Shlyapnikov}},\ }%
  \bibfield{journal}{%
  \Doi{10.1103/PhysRevA.60.R2665}{\bibinfo {journal} {Phys. Rev. A}}\ }%
  \textbf{\bibinfo {volume} {60}},\ \bibinfo {pages} {R2665} (\bibinfo {year}
  {1999})%
  \bibAnnoteFile{NoStop}{PhysRevA.60.R2665}%
\bibitem{PhysRevA.64.063602}%
  \BibitemOpen
  \bibfield{author}{%
  \bibinfo {author} {\bibfnamefont{J.-P.}\ \bibnamefont{Martikainen}}, \bibinfo
  {author} {\bibfnamefont{K.-A.}\ \bibnamefont{Suominen}}, \bibinfo {author}
  {\bibfnamefont{L.}~\bibnamefont{Santos}}, \bibinfo {author}
  {\bibfnamefont{T.}~\bibnamefont{Schulte}},\ and\ \bibinfo {author}
  {\bibfnamefont{A.}~\bibnamefont{Sanpera}},\ }%
  \bibfield{journal}{%
  \Doi{10.1103/PhysRevA.64.063602}{\bibinfo {journal} {Phys. Rev. A}}\ }%
  \textbf{\bibinfo {volume} {64}},\ \bibinfo {pages} {063602} (\bibinfo {year}
  {2001})%
  \bibAnnoteFile{NoStop}{PhysRevA.64.063602}%
\bibitem{ZacharyDutton07272001}%
  \BibitemOpen
  \bibfield{author}{%
  \bibinfo {author} {\bibfnamefont{Z.}~\bibnamefont{Dutton}}, \bibinfo {author}
  {\bibfnamefont{M.}~\bibnamefont{Budde}}, \bibinfo {author}
  {\bibfnamefont{C.}~\bibnamefont{Slowe}},\ and\ \bibinfo {author}
  {\bibfnamefont{L.~V.}\ \bibnamefont{Hau}},\ }%
  \bibfield{journal}{%
  \Doi{10.1126/science.1062527}{\bibinfo {journal} {Science}}\ }%
  \textbf{\bibinfo {volume} {293}},\ \bibinfo {pages} {663} (\bibinfo {year}
  {2001})%
  \bibAnnoteFile{NoStop}{ZacharyDutton07272001}%
\bibitem{PhysRevLett.86.2926}%
  \BibitemOpen
  \bibfield{author}{%
  \bibinfo {author} {\bibfnamefont{B.~P.}\ \bibnamefont{Anderson}}, \bibinfo
  {author} {\bibfnamefont{P.~C.}\ \bibnamefont{Haljan}}, \bibinfo {author}
  {\bibfnamefont{C.~A.}\ \bibnamefont{Regal}}, \bibinfo {author}
  {\bibfnamefont{D.~L.}\ \bibnamefont{Feder}}, \bibinfo {author}
  {\bibfnamefont{L.~A.}\ \bibnamefont{Collins}}, \bibinfo {author}
  {\bibfnamefont{C.~W.}\ \bibnamefont{Clark}},\ and\ \bibinfo {author}
  {\bibfnamefont{E.~A.}\ \bibnamefont{Cornell}},\ }%
  \bibfield{journal}{%
  \Doi{10.1103/PhysRevLett.86.2926}{\bibinfo {journal} {Phys. Rev. Lett.}}\ }%
  \textbf{\bibinfo {volume} {86}},\ \bibinfo {pages} {2926} (\bibinfo {year}
  {2001})%
  \bibAnnoteFile{NoStop}{PhysRevLett.86.2926}%
\bibitem{PhysRevA.87.043601}%
  \BibitemOpen
  \bibfield{author}{%
  \bibinfo {author} {\bibfnamefont{L.~A.}\ \bibnamefont{Toikka}}\ and\ \bibinfo
  {author} {\bibfnamefont{K.-A.}\ \bibnamefont{Suominen}},\ }%
  \bibfield{journal}{%
  \Doi{10.1103/PhysRevA.87.043601}{\bibinfo {journal} {Phys. Rev. A}}\ }%
  \textbf{\bibinfo {volume} {87}},\ \bibinfo {pages} {043601} (\bibinfo {year}
  {2013})%
  \bibAnnoteFile{NoStop}{PhysRevA.87.043601}%
\bibitem{PhysRevLett.83.5198}%
  \BibitemOpen
  \bibfield{author}{%
  \bibinfo {author} {\bibfnamefont{S.}~\bibnamefont{Burger}}, \bibinfo {author}
  {\bibfnamefont{K.}~\bibnamefont{Bongs}}, \bibinfo {author}
  {\bibfnamefont{S.}~\bibnamefont{Dettmer}}, \bibinfo {author}
  {\bibfnamefont{W.}~\bibnamefont{Ertmer}}, \bibinfo {author}
  {\bibfnamefont{K.}~\bibnamefont{Sengstock}}, \bibinfo {author}
  {\bibfnamefont{A.}~\bibnamefont{Sanpera}}, \bibinfo {author}
  {\bibfnamefont{G.~V.}\ \bibnamefont{Shlyapnikov}},\ and\ \bibinfo {author}
  {\bibfnamefont{M.}~\bibnamefont{Lewenstein}},\ }%
  \bibfield{journal}{%
  \Doi{10.1103/PhysRevLett.83.5198}{\bibinfo {journal} {Phys. Rev. Lett.}}\ }%
  \textbf{\bibinfo {volume} {83}},\ \bibinfo {pages} {5198} (\bibinfo {year}
  {1999})%
  \bibAnnoteFile{NoStop}{PhysRevLett.83.5198}%
\bibitem{J.Denschlag01072000}%
  \BibitemOpen
  \bibfield{author}{%
  \bibinfo {author} {\bibfnamefont{J.}~\bibnamefont{Denschlag}}, \bibinfo
  {author} {\bibfnamefont{J.~E.}\ \bibnamefont{Simsarian}}, \bibinfo {author}
  {\bibfnamefont{D.~L.}\ \bibnamefont{Feder}}, \bibinfo {author}
  {\bibfnamefont{C.~W.}\ \bibnamefont{Clark}}, \bibinfo {author}
  {\bibfnamefont{L.~A.}\ \bibnamefont{Collins}}, \bibinfo {author}
  {\bibfnamefont{J.}~\bibnamefont{Cubizolles}}, \bibinfo {author}
  {\bibfnamefont{L.}~\bibnamefont{Deng}}, \bibinfo {author}
  {\bibfnamefont{E.~W.}\ \bibnamefont{Hagley}}, \bibinfo {author}
  {\bibfnamefont{K.}~\bibnamefont{Helmerson}}, \bibinfo {author}
  {\bibfnamefont{W.~P.}\ \bibnamefont{Reinhardt}}, \bibinfo {author}
  {\bibfnamefont{S.~L.}\ \bibnamefont{Rolston}}, \bibinfo {author}
  {\bibfnamefont{B.~I.}\ \bibnamefont{Schneider}},\ and\ \bibinfo {author}
  {\bibfnamefont{W.~D.}\ \bibnamefont{Phillips}},\ }%
  \bibfield{journal}{%
  \Doi{10.1126/science.287.5450.97}{\bibinfo {journal} {Science}}\ }%
  \textbf{\bibinfo {volume} {287}},\ \bibinfo {pages} {97} (\bibinfo {year}
  {2000})%
  \bibAnnoteFile{NoStop}{J.Denschlag01072000}%
\bibitem{Becker08}%
  \BibitemOpen
  \bibfield{author}{%
  \bibinfo {author} {\bibfnamefont{C.}~\bibnamefont{Becker}}, \bibinfo {author}
  {\bibfnamefont{S.}~\bibnamefont{Stellmer}}, \bibinfo {author}
  {\bibfnamefont{P.}~\bibnamefont{Soltan-Panahi}}, \bibinfo {author}
  {\bibfnamefont{S.}~\bibnamefont{D\"{o}rscher}}, \bibinfo {author}
  {\bibfnamefont{M.}~\bibnamefont{Baumert}}, \bibinfo {author}
  {\bibfnamefont{E.-M.}\ \bibnamefont{Richter}}, \bibinfo {author}
  {\bibfnamefont{J.}~\bibnamefont{Kronj\"{a}ger}}, \bibinfo {author}
  {\bibfnamefont{K.}~\bibnamefont{Bongs}},\ and\ \bibinfo {author}
  {\bibfnamefont{K.}~\bibnamefont{Sengstock}},\ }%
  \bibfield{journal}{%
  \Doi{10.1038/nphys962}{\bibinfo {journal} {Nature {P}hysics}}\ }%
  \textbf{\bibinfo {volume} {4}},\ \bibinfo {pages} {496} (\bibinfo {year}
  {2008})%
  \bibAnnoteFile{NoStop}{Becker08}%
\bibitem{0953-4075-30-22-001}%
  \BibitemOpen
  \bibfield{author}{%
  \bibinfo {author} {\bibfnamefont{W.~P.}\ \bibnamefont{Reinhardt}}\ and\
  \bibinfo {author} {\bibfnamefont{C.~W.}\ \bibnamefont{Clark}},\ }%
  \bibfield{journal}{%
  \Doi{10.1088/0953-4075/30/22/001}{\bibinfo {journal} {J. Phys. B}}\ }%
  \textbf{\bibinfo {volume} {30}},\ \bibinfo {pages} {L785} (\bibinfo {year}
  {1997})%
  \bibAnnoteFile{NoStop}{0953-4075-30-22-001}%
\bibitem{Shomroni09}%
  \BibitemOpen
  \bibfield{author}{%
  \bibinfo {author} {\bibfnamefont{I.}~\bibnamefont{Shomroni}}, \bibinfo
  {author} {\bibfnamefont{E.}~\bibnamefont{Lahoud}}, \bibinfo {author}
  {\bibfnamefont{S.}~\bibnamefont{Levy}},\ and\ \bibinfo {author}
  {\bibfnamefont{J.}~\bibnamefont{Steinhauer}},\ }%
  \bibfield{journal}{%
  \Doi{10.1038/nphys1177}{\bibinfo {journal} {Nature {P}hysics}}\ }%
  \textbf{\bibinfo {volume} {5}},\ \bibinfo {pages} {193} (\bibinfo {year}
  {2009})%
  \bibAnnoteFile{NoStop}{Shomroni09}%
\bibitem{ths2012}%
  \BibitemOpen
  \bibfield{author}{%
  \bibinfo {author} {\bibfnamefont{L.~A.}\ \bibnamefont{Toikka}}, \bibinfo
  {author} {\bibfnamefont{J.}~\bibnamefont{Hietarinta}},\ and\ \bibinfo
  {author} {\bibfnamefont{K.-A.}\ \bibnamefont{Suominen}},\ }%
  \bibfield{journal}{%
  \Doi{10.1088/1751-8113/45/48/485203}{\bibinfo {journal} {J. Phys. A}}\ }%
  \textbf{\bibinfo {volume} {45}},\ \bibinfo {pages} {485203} (\bibinfo {year}
  {2012})%
  \bibAnnoteFile{NoStop}{ths2012}%
\bibitem{Andrews31011997}%
  \BibitemOpen
  \bibfield{author}{%
  \bibinfo {author} {\bibfnamefont{M.~R.}\ \bibnamefont{Andrews}}, \bibinfo
  {author} {\bibfnamefont{C.~G.}\ \bibnamefont{Townsend}}, \bibinfo {author}
  {\bibfnamefont{H.-J.}\ \bibnamefont{Miesner}}, \bibinfo {author}
  {\bibfnamefont{D.~S.}\ \bibnamefont{Durfee}}, \bibinfo {author}
  {\bibfnamefont{D.~M.}\ \bibnamefont{Kurn}},\ and\ \bibinfo {author}
  {\bibfnamefont{W.}~\bibnamefont{Ketterle}},\ }%
  \bibfield{journal}{%
  \Doi{10.1126/science.275.5300.637}{\bibinfo {journal} {Science}}\ }%
  \textbf{\bibinfo {volume} {275}},\ \bibinfo {pages} {637} (\bibinfo {year}
  {1997})%
  \bibAnnoteFile{NoStop}{Andrews31011997}%
\bibitem{0953-4075-31-8-001}%
  \BibitemOpen
  \bibfield{author}{%
  \bibinfo {author} {\bibfnamefont{T.~F.}\ \bibnamefont{Scott}}, \bibinfo
  {author} {\bibfnamefont{R.~J.}\ \bibnamefont{Ballagh}},\ and\ \bibinfo
  {author} {\bibfnamefont{K.}~\bibnamefont{Burnett}},\ }%
  \bibfield{journal}{%
  \Doi{10.1088/0953-4075/31/8/001}{\bibinfo {journal} {J. Phys. B}}\ }%
  \textbf{\bibinfo {volume} {31}},\ \bibinfo {pages} {L329} (\bibinfo {year}
  {1998})%
  \bibAnnoteFile{NoStop}{0953-4075-31-8-001}%
\bibitem{PhysRevA.74.043613}%
  \BibitemOpen
  \bibfield{author}{%
  \bibinfo {author} {\bibfnamefont{L.~D.}\ \bibnamefont{Carr}}\ and\ \bibinfo
  {author} {\bibfnamefont{C.~W.}\ \bibnamefont{Clark}},\ }%
  \bibfield{journal}{%
  \Doi{10.1103/PhysRevA.74.043613}{\bibinfo {journal} {Phys. Rev. A}}\ }%
  \textbf{\bibinfo {volume} {74}},\ \bibinfo {pages} {043613} (\bibinfo {year}
  {2006})%
  \bibAnnoteFile{NoStop}{PhysRevA.74.043613}%
\bibitem{Garraway1998}%
  \BibitemOpen
  \bibfield{author}{%
  \bibinfo {author} {\bibfnamefont{B.~M.}\ \bibnamefont{Garraway}}\ and\
  \bibinfo {author} {\bibfnamefont{K.-A.}\ \bibnamefont{Suominen}},\ }%
  \bibfield{journal}{%
  \Doi{10.1103/PhysRevLett.80.932}{\bibinfo {journal} {Phys. Rev. Lett.}}\ }%
  \textbf{\bibinfo {volume} {80}},\ \bibinfo {pages} {932} (\bibinfo {year}
  {1998})%
  \bibAnnoteFile{NoStop}{Garraway1998}%
\bibitem{Rodriguez2000}%
  \BibitemOpen
  \bibfield{author}{%
  \bibinfo {author} {\bibfnamefont{M.}~\bibnamefont{Rodriguez}}, \bibinfo
  {author} {\bibfnamefont{K.-A.}\ \bibnamefont{Suominen}},\ and\ \bibinfo
  {author} {\bibfnamefont{B.~M.}\ \bibnamefont{Garraway}},\ }%
  \bibfield{journal}{%
  \Doi{10.1103/PhysRevA.62.053413}{\bibinfo {journal} {Phys. Rev. A}}\ }%
  \textbf{\bibinfo {volume} {62}},\ \bibinfo {pages} {053413} (\bibinfo {year}
  {2000})%
  \bibAnnoteFile{NoStop}{Rodriguez2000}%
\bibitem{PROP:PROP200310015}%
  \BibitemOpen
  \bibfield{author}{%
  \bibinfo {author} {\bibfnamefont{B.}~\bibnamefont{Garraway}}\ and\ \bibinfo
  {author} {\bibfnamefont{K.-A.}\ \bibnamefont{Suominen}},\ }%
  \bibfield{journal}{%
  \Doi{10.1002/prop.200310015}{\bibinfo {journal} {Fortschr. Phys.}}\ }%
  \textbf{\bibinfo {volume} {51}},\ \bibinfo {pages} {128} (\bibinfo {year}
  {2003})%
  \bibAnnoteFile{NoStop}{PROP:PROP200310015}%
\bibitem{vitanov2001laser}%
  \BibitemOpen
  \bibfield{author}{%
  \bibinfo {author} {\bibfnamefont{N.~V.}\ \bibnamefont{Vitanov}}, \bibinfo
  {author} {\bibfnamefont{T.}~\bibnamefont{Halfmann}}, \bibinfo {author}
  {\bibfnamefont{B.~W.}\ \bibnamefont{Shore}},\ and\ \bibinfo {author}
  {\bibfnamefont{K.}~\bibnamefont{Bergmann}},\ }%
  \bibfield{journal}{%
  \Doi{10.1146/annurev.physchem.52.1.763}{\bibinfo {journal} {Annu. Rev. Phys.
  Chem.}}\ }%
  \textbf{\bibinfo {volume} {52}},\ \bibinfo {pages} {763} (\bibinfo {year}
  {2001})%
  \bibAnnoteFile{NoStop}{vitanov2001laser}%
\bibitem{Ljp47318}%
  \BibitemOpen
  \bibfield{author}{%
  \bibinfo {author} {\bibfnamefont{G.}~\bibnamefont{Juzeli\={u}nas}}, \bibinfo
  {author} {\bibfnamefont{J.}~\bibnamefont{Ruseckas}}, \bibinfo {author}
  {\bibfnamefont{P.}~\bibnamefont{\"{O}hberg}},\ and\ \bibinfo {author}
  {\bibfnamefont{M.}~\bibnamefont{Fleischhauer}},\ }%
  \bibfield{journal}{%
  \bibinfo {journal} {Lithuanian J. Phys.}\ }%
  \textbf{\bibinfo {volume} {47}},\ \bibinfo {pages} {351} (\bibinfo {year}
  {2007})%
  \bibAnnoteFile{NoStop}{Ljp47318}%
\bibitem{PhysRevLett.80.2972}%
  \BibitemOpen
  \bibfield{author}{%
  \bibinfo {author} {\bibfnamefont{R.}~\bibnamefont{Dum}}, \bibinfo {author}
  {\bibfnamefont{J.~I.}\ \bibnamefont{Cirac}}, \bibinfo {author}
  {\bibfnamefont{M.}~\bibnamefont{Lewenstein}},\ and\ \bibinfo {author}
  {\bibfnamefont{P.}~\bibnamefont{Zoller}},\ }%
  \bibfield{journal}{%
  \Doi{10.1103/PhysRevLett.80.2972}{\bibinfo {journal} {Phys. Rev. Lett.}}\ }%
  \textbf{\bibinfo {volume} {80}},\ \bibinfo {pages} {2972} (\bibinfo {year}
  {1998})%
  \bibAnnoteFile{NoStop}{PhysRevLett.80.2972}%
\bibitem{Harkonen2006}%
  \BibitemOpen
  \bibfield{author}{%
  \bibinfo {author} {\bibfnamefont{K.}~\bibnamefont{H\"ark\"onen}}, \bibinfo
  {author} {\bibfnamefont{O.}~\bibnamefont{K\"arki}},\ and\ \bibinfo {author}
  {\bibfnamefont{K.-A.}\ \bibnamefont{Suominen}},\ }%
  \bibfield{journal}{%
  \Doi{10.1103/PhysRevA.74.043404}{\bibinfo {journal} {Phys. Rev. A}}\ }%
  \textbf{\bibinfo {volume} {74}},\ \bibinfo {pages} {043404} (\bibinfo {year}
  {2006})%
  \bibAnnoteFile{NoStop}{Harkonen2006}%
\bibitem{Toikka2013}%
  \BibitemOpen
  \bibfield{author}{%
  \bibinfo {author} {\bibfnamefont{L.~A.}\ \bibnamefont{Toikka}},\ }%
  \bibfield{journal}{%
  \bibinfo {journal} {unpublished}}%
   (\bibinfo {year} {2013})%
  \bibAnnoteFile{NoStop}{Toikka2013}%
\bibitem{PhysRevLett.102.090402}%
  \BibitemOpen
  \bibfield{author}{%
  \bibinfo {author} {\bibfnamefont{S.~E.}\ \bibnamefont{Pollack}}, \bibinfo
  {author} {\bibfnamefont{D.}~\bibnamefont{Dries}}, \bibinfo {author}
  {\bibfnamefont{M.}~\bibnamefont{Junker}}, \bibinfo {author}
  {\bibfnamefont{Y.~P.}\ \bibnamefont{Chen}}, \bibinfo {author}
  {\bibfnamefont{T.~A.}\ \bibnamefont{Corcovilos}},\ and\ \bibinfo {author}
  {\bibfnamefont{R.~G.}\ \bibnamefont{Hulet}},\ }%
  \bibfield{journal}{%
  \Doi{10.1103/PhysRevLett.102.090402}{\bibinfo {journal} {Phys. Rev. Lett.}}\
  }%
  \textbf{\bibinfo {volume} {102}},\ \bibinfo {pages} {090402} (\bibinfo {year}
  {2009})%
  \bibAnnoteFile{NoStop}{PhysRevLett.102.090402}%
\bibitem{SM_Video}%
  \BibitemOpen
  \bibinfo {note} {See {S}upplemental {M}aterial at [url] for video}%
  \bibAnnoteFile{NoStop}{SM_Video}%
\bibitem{gerry2005introductory}%
  \BibitemOpen
  \bibfield{author}{%
  \bibinfo {author} {\bibfnamefont{C.}~\bibnamefont{Gerry}}\ and\ \bibinfo
  {author} {\bibfnamefont{P.}~\bibnamefont{Knight}},\ }%
  \emph{\bibinfo {title} {{I}ntroductory {Q}uantum {O}ptics}}\ (\bibinfo
  {publisher} {Cambridge University Press},\ \bibinfo {year} {2005})%
  \bibAnnoteFile{NoStop}{gerry2005introductory}%
\bibitem{PhysRevLett.104.075301}%
  \BibitemOpen
  \bibfield{author}{%
  \bibinfo {author} {\bibfnamefont{A.~C.}\ \bibnamefont{White}}, \bibinfo
  {author} {\bibfnamefont{C.~F.}\ \bibnamefont{Barenghi}}, \bibinfo {author}
  {\bibfnamefont{N.~P.}\ \bibnamefont{Proukakis}}, \bibinfo {author}
  {\bibfnamefont{A.~J.}\ \bibnamefont{Youd}},\ and\ \bibinfo {author}
  {\bibfnamefont{D.~H.}\ \bibnamefont{Wacks}},\ }%
  \bibfield{journal}{%
  \Doi{10.1103/PhysRevLett.104.075301}{\bibinfo {journal} {Phys. Rev. Lett.}}\
  }%
  \textbf{\bibinfo {volume} {104}},\ \bibinfo {pages} {075301} (\bibinfo {year}
  {2010})%
  \bibAnnoteFile{NoStop}{PhysRevLett.104.075301}%
\bibitem{PhysRevE.84.067301}%
  \BibitemOpen
  \bibfield{author}{%
  \bibinfo {author} {\bibfnamefont{A.~W.}\ \bibnamefont{Baggaley}}\ and\
  \bibinfo {author} {\bibfnamefont{C.~F.}\ \bibnamefont{Barenghi}},\ }%
  \bibfield{journal}{%
  \Doi{10.1103/PhysRevE.84.067301}{\bibinfo {journal} {Phys. Rev. E}}\ }%
  \textbf{\bibinfo {volume} {84}},\ \bibinfo {pages} {067301} (\bibinfo {year}
  {2011})%
  \bibAnnoteFile{NoStop}{PhysRevE.84.067301}%
\bibitem{PhysRevLett.101.154501}%
  \BibitemOpen
  \bibfield{author}{%
  \bibinfo {author} {\bibfnamefont{M.~S.}\ \bibnamefont{Paoletti}}, \bibinfo
  {author} {\bibfnamefont{M.~E.}\ \bibnamefont{Fisher}}, \bibinfo {author}
  {\bibfnamefont{K.~R.}\ \bibnamefont{Sreenivasan}},\ and\ \bibinfo {author}
  {\bibfnamefont{D.~P.}\ \bibnamefont{Lathrop}},\ }%
  \bibfield{journal}{%
  \Doi{10.1103/PhysRevLett.101.154501}{\bibinfo {journal} {Phys. Rev. Lett.}}\
  }%
  \textbf{\bibinfo {volume} {101}},\ \bibinfo {pages} {154501} (\bibinfo {year}
  {2008})%
  \bibAnnoteFile{NoStop}{PhysRevLett.101.154501}%
\bibitem{PhysRevLett.80.3899}%
  \BibitemOpen
  \bibfield{author}{%
  \bibinfo {author} {\bibfnamefont{R.}~\bibnamefont{Dum}}, \bibinfo {author}
  {\bibfnamefont{A.}~\bibnamefont{Sanpera}}, \bibinfo {author}
  {\bibfnamefont{K.-A.}\ \bibnamefont{Suominen}}, \bibinfo {author}
  {\bibfnamefont{M.}~\bibnamefont{Brewczyk}}, \bibinfo {author}
  {\bibfnamefont{M.}~\bibnamefont{Ku\a'{s}}}, \bibinfo {author}
  {\bibfnamefont{K.}~\bibnamefont{Rz\textpolhook{a}\.{z}ewski}},\ and\ \bibinfo
  {author} {\bibfnamefont{M.}~\bibnamefont{Lewenstein}},\ }%
  \bibfield{journal}{%
  \Doi{10.1103/PhysRevLett.80.3899}{\bibinfo {journal} {Phys. Rev. Lett.}}\ }%
  \textbf{\bibinfo {volume} {80}},\ \bibinfo {pages} {3899} (\bibinfo {year}
  {1998})%
  \bibAnnoteFile{NoStop}{PhysRevLett.80.3899}%
\end{thebibliography}%
\end{document}